\documentclass[a4paper,unsortedaddress,twocolumn,superscriptaddress,aip,jcp]{revtex4}

\usepackage[english]{babel}
\usepackage{natbib}
\usepackage{graphicx}
\usepackage{dcolumn}
\usepackage{bm}
\usepackage{epstopdf}
\usepackage{pstool}
\usepackage[T1]{fontenc}
\usepackage{lmodern}
\usepackage[colorlinks=true, pdfstartview=FitV, linkcolor=blue, 
            citecolor=blue, urlcolor=blue]{hyperref}
\usepackage{bookmark}
\usepackage{epsfig}
\usepackage{microtype}
\usepackage[caption=false]{subfig}
\usepackage{braket}
\usepackage[font=small,format=plain,labelfont=bf,up,textfont=small,up,justification=centerlast,position=below]{caption}
\usepackage{amsmath,amsfonts,amssymb}

\newcommand{\dd}{{\rm d}}
\newcommand{\be}{\begin{equation}}
\newcommand{\ee}{\end{equation}}
\newcommand{\ba}{\begin{eqnarray}}
\newcommand{\ea}{\end{eqnarray}}

\EndPreamble

\begin{document}
\graphicspath{{figures/}}

\title{Efficient and accurate solver of the three-dimensional screened and unscreened Poisson's equation with generic boundary conditions}

\author{Alessandro \surname{Cerioni}}
\email{alessandro.cerioni@esrf.fr}
\affiliation{European Synchrotron Radiation Facility, 6 rue Horowitz, BP 220, 38043 Grenoble Cedex 9, France}

\author{Luigi \surname{Genovese}}
\email{luigi.genovese@cea.fr}
\affiliation{Laboratoire de simulation atomistique (L\_Sim), SP2M, UMR-E CEA / UJF-Grenoble 1, INAC, Grenoble, F-38054, France}

\author{Alessandro \surname{Mirone}}
\author{Vicente Armando \surname{Sole}}  
\affiliation{European Synchrotron Radiation Facility, 6 rue Horowitz, BP 220, 38043 Grenoble Cedex 9, France}

\begin{abstract}
We present an explicit solver of the three-dimensional screened and unscreened Poisson's equation which combines accuracy, computational efficiency and versatility. The solver, based on a mixed plane-wave / interpolating scaling function representation, can deal with any kind of periodicity (along one, two, or three spatial axes) as well as with fully isolated boundary conditions. It can seamlessly accommodate a finite screening length, non-orthorhombic lattices and charged systems. This approach is particularly advantageous because convergence is attained by simply refining the real space grid, namely without any adjustable parameter. At the same time, the numerical method features $\mathcal{O}(N\log N)$ scaling of the computational cost ($N$ being the number of grid points) very much like plane-wave methods. The methodology, validated on model systems, is tailored for leading-edge computer simulations of materials (including \emph{ab initio} electronic structure computations), but it might as well be beneficial for other research domains.
\end{abstract}

\maketitle

\section{Introduction}
Poisson's equation (screened or not) is involved in a large variety of problems in physics and chemistry as well as in engineering. Therefore, there is a quite strong motivation for developing efficient and accurate solving methods. 

As far as electrostatics is concerned, the three-dimensional screened Poisson's equation is written as follows (in Gaussian units):
\be\label{eq:screened_Poisson}
(\nabla^2-\mu_0^2) V(x,y,z) = -4\pi \rho(x,y,z),
\ee
where $\rho(x,y,z)$ represents a continuous electric charge distribution (the input of the problem at hand), $V(x,y,z)$ is the electrostatic potential (the output), and $\mu_0$ represents the reciprocal screening length as defined, for instance, in the Debye-H\"uckel or Thomas-Fermi approximations. In the special case $\mu_0=0$, Eq.\ \eqref{eq:screened_Poisson} reduces to the usual Poisson's equation.

Any method aiming at providing a solution to Eq.\ \eqref{eq:screened_Poisson} has to deal with boundary conditions (BC), which in general
can be either periodic or free (otherwise referred to as ``isolated'' or ``open''') along each of the three directions $x,y,z$. In the case of fully periodic BC, the most natural (and efficient) approach to the problem is that of the reciprocal space treatment. It amounts to expanding both the density and the potential as superpositions of plane waves (Fourier series), following which Eq.\ \eqref{eq:screened_Poisson} becomes algebraic in the Fourier components of $\rho$ and $V$. This equation is readily solved and the result is finally transformed back into real space. Forward and backward transformations are carried out via Fast Fourier Transform (FFT), hence the overall computational scaling of the method with respect to the number $N$ of grid points is a rather appealing $\mathcal{O}(N\log N)$. 

Owing to the above mentioned advantages, many attempts have been made to tackle also free BC or mixed free/periodic BC within a nominally fully periodic framework. In the simplest implementations, the simulation box is artificially enlarged by vacuum padding, such as to suppress the spurious Coulombian interaction among periodic replicas (``super-cell'' approximation). Yet, as a result of the the long-range nature of the Coulomb interaction, there are situations in which the simulation box ought to be made unfeasibly large, in particular for charge distributions exhibiting significant multipolar terms. Moreover, a non-zero net charge in the primitive cell would yield a divergent total electrostatic energy when infinitely replicated along every direction, unless a compensating artificial uniformly charged background (``jellium'') is introduced, see \emph{e.g.}\ Refs.\ \cite{0022-3719-18-5-005} and \cite{PhysRevB.51.4014}. Another option which has been put forward consists in cutting off the long tail of the Coulomb interaction beyond a spherical region in real space, the radius of which is adequately chosen. Correspondingly, the reciprocal space components of the bare Coulomb potential are multiplied by screening functions, known analytically for all types of periodicity \cite{PhysRevB.56.14972,PhysRevB.73.205119}. In a series of papers \cite{martyna:clusters, minary:surfaces, minary:wires},  the screening function formalism was combined with the explicit break-up of the short- and long-range components of the Coulomb interaction, as also done in the context of (smooth particle-mesh) Ewald summation techniques, whereas in Refs.\ \cite{PhysRevB.77.115139, PhysRevB.84.159910} the errors induced by the periodic images are alleviated by introducing a corrective potential.

Although we acknowledge that much remarkable work has been done on the subject, providing a thorough review would go beyond the scope of the present study. We therefore refer the reader to the original literature and proceed by presenting our approach, which differs from those mentioned above in that convergence is attained with no adjustable parameter and thus it aims at being fully generic.  

This paper builds upon Refs.\ \cite{genovese:054704,genovese:074105}, where a novel method for solving the unscreened Poisson's equation with free and surface-like BC was first presented. Such a method is direct (rather than iterative) in that the solution along the isolated directions is found in its integral form by using  the Green's function method. For instance, in the case of a fully isolated system (or ``cluster-like''),
\be\label{eq:convolution_integral}
V(\vec{r}) = 4\pi \int \dd^3\vec{r}'\, G(\mu_0; |\vec{r}-\vec{r'}|) \rho(\vec{r}'),
\ee
where $\vec{r} = (x,y,z)$. Homogeneous Dirichlet BC ($V = 0$ at $|\vec{r}|\rightarrow \infty$) along the isolated directions are explicitly enforced by the selection of the Green's function.

The method has been in use for a few years in a number of \textit{ab-initio} codes, namely \textsc{ABINIT} \cite{ABINIT1,Gonze20092582}, \textsc{BigDFT}  \cite{genovese:014109,Genovese2011149}, \textsc{CP2K} \cite{VandeVondele2005103} (see also Ref.\ \cite{doi:10.1021/jz2014852} for a recent application thereof), \textsc{Octopus} \cite{Marques200360,PSSB:PSSB200642067} and has proven to be highly efficient and accurate in every application attempted to date. It is based on a mixed plane-wave / interpolating scaling function (ISF) representation of  $\rho$ and $V$ which allows to model any sort of periodicity in the most natural, clean and mathematically rigorous way.  Clearly, periodic (isolated) directions are represented in terms of plane waves (interpolating scaling functions). ISFs - arising in the wavelet theory \cite{daub,sgbook} -  enjoy several properties which make them superior to other basis sets. For instance, the representation in terms of $m$-th order ISFs make the first $m$ moments of the continuous and discrete charge distributions coincide \cite{genovese:074105}. As a consequence the representation is definitely faithful (other than handy), since the different moments of the charge distribution capture the major features of the potential. Moreover, ISFs are genuinely localized due to their compact support (the length of which is equal to $2m$) and endowed with the so-called ``refinement relations'' which easily allow to switch from a representation on a grid with spacing $h$ to a doubly refined grid with spacing $h/2$.

We have extended the previous implementation to account also for screening, for the case of periodicity along only one direction (``wire-like BC'') and for non-orthorhombic cells, this investigation providing a detailed account of such improvements. The inclusion of such new functionalities is  motivated by the strong theoretical, experimental and technological interest in the characterization of nanostructured materials (among which polar nanorods, see \emph{e.g.}\ Ref.\ \cite{1742-6596-367-1-012002} and references therein), since solving Poisson's equation is only one of the many steps involved in state-of-the-art computer simulations and is repeated several times. Moreover, in the context of Kohn-Sham (KS) density functional theory (DFT) and extensions thereof, there are quantities which are computed via convolution integrals very similar to that in Eq.\ \eqref{eq:convolution_integral}: for instance, the exact exchange term arising within those generalizations of KS-DFT employing orbital-dependent or hybrid functionals (see \cite{doi:10.1021/ct2009363} and references therein), or the coupling-matrix in time-dependent DFT \cite{Natarajan201229}. In this respect, the electrostatic problem of concern here provides the paradigm for many other computations, even well beyond the scope of electrostatics. 

Regarding the possibility of accounting also for screening, we note that this novel feature might for example be used to solve the Schr\"odinger equation iteratively (see \emph{e.g.}\ Ref.\ \cite{harrison:11587}). In fact, one can exploit the formal analogy between Eq.\ \eqref{eq:screened_Poisson} and the Schr\"odinger equation which becomes apparent if the latter is written in the following fashion:
\be
\left(\frac{\hbar^2}{2m}\nabla^2-|E|\right)\ket{\psi} = V\ket{\psi},
\ee
where $\ket{\psi}$ represents a bound eigenstate with negative energy ($E<0$).

The next sections are structured as follows: we first present our solution method for free, wire-like and surface-like BC. We then discuss the accuracy of the proposed solver by reporting a collection of numerical benchmarks. We conclude by highlighting the benefits of using the present methodology.

\section{Free Boundary Conditions}
\label{sec:FBC}
In the case of free BC, the Green's function which has to be plugged into Eq.\ \eqref{eq:convolution_integral} is 
\be
G(\mu_0;r) = \frac{e^{-\mu_0 r}}{4\pi r},
\ee
since 
\be
\left(\frac{\partial^2}{\partial r^2}+\frac{2}{r}\frac{\partial}{\partial r}-\mu_0^2\right)G(\mu_0; r) = -\delta^{(3)} (r),
\ee
where  $r =|\vec{r}|$. Along the same lines as Ref.\ \cite{genovese:074105}, both the charge distribution and the electrostatic potential are expanded in terms of ISFs, here denoted by $\phi$:
\begin{widetext}
\be\label{eq:grid_density}
\rho(x,y,z) = \sum_{j_x = 0}^{N_x}\sum_{j_y = 0}^{N_y}\sum_{j_z = 0}^{N_z}\rho_{j_x, j_y, j_z}\,\phi\left(\frac{x}{h_x}-j_x \right)\phi\left(\frac{y}{h_y}-j_y \right)\phi\left(\frac{z}{h_z}-j_z \right)\,,
\ee
\end{widetext}
where $h_{\{x,y,z\}}$ and $N_{\{x,y,z\}}$ represent the (uniform) grid spacing and the number of grid points along each direction, respectively. Since the function $\phi$ is, by construction, such that $\phi(j) =\delta_{j,0}$, $\forall j \in \mathbb{Z}$, the expansion coefficients $\rho_{j_x, j_y, j_z}$ are readily found to be:
\be\label{eq:density_on_grid}
\rho_{j_x, j_y, j_z} = \rho(h_x j_x, h_y j_y, h_z j_z).
\ee
In other words, the expansion coefficients coincide exactly with the values of $\rho(x,y,z)$ on a uniform grid. In this respect, the ISF representation appears genuinely tailored for numerical studies, where the whole available information reduces to knowing the values at the grid points. Clearly, the grid spacing has to be chosen adequately, depending on the typical spatial scales over which the density distribution exhibits significant variations. The underlying mathematics then assures that the moments built upon the discrete charge distribution coincide with those of the continuous charge distribution up to order $m-1$, where $m$ is the order of the ISF (see Ref.\  \cite{genovese:074105} for the proof):
\be
\sum_{i, j, k} i^{l_1}\,j^{l_2}\,k^{l_3}\; \rho_{i,j,k} = \int \dd^3 \vec{r}\; x^{l_1} y^{l_2} z^{l_3} \rho(\vec{r})
\ee
if $0\leq l_1, l_2, l_3<m$. A representation analogous to that in Eq.\ \eqref{eq:grid_density} can also be given for the potential $V$, where 
\be\label{eq:grid_potential}
V_{j_x,j_y,j_z} = V(h_x j_x, h_y j_y, h_z j_z)
\ee
replaces $\rho_{j_x, j_y, j_z}$. As a consequence of the chosen representation, the convolution integral in Eq.\ \eqref{eq:convolution_integral} is more conveniently expressed in Cartesian coordinates, and upon plugging Eq.\ \eqref{eq:grid_density} into Eq.\ \eqref{eq:convolution_integral}  we  obtain the following equation in discrete form:
\begin{widetext}
\be
V_{j_x,j_y,j_z}  = 4\pi h_x h_y h_z  \sum_{j'_x = 0}^{N_x}\sum_{j'_y = 0}^{N_y}\sum_{j'_z = 0}^{N_z}K (j_x- j_x', j_y- j_y',j_z- j'_z;\mu_0)\,\rho_{j'_x, j'_y, j'_z},
\ee
where 
\be\label{eq:kernel_FBC}
K(j_x,j_y,j_z;\mu) =  \int \dd u \dd v \dd w \,G(\mu;\sqrt{[h_x  (j_x-u)]^2+[h_y (j_y-v)]^2+[h_z( j_z-w)]^2}) \, \phi(u)\phi(v)\phi(w).
\ee
is the convolution kernel. 
\end{widetext}

The numerical evaluation of Eq.\ \eqref{eq:kernel_FBC} would be too onerous if performed directly. The computational effort can be drastically reduced by expressing the Green's function as a linear combination of Gaussian functions, as the integral would become separable along $x,y,z$ and the resulting 1D integrals can be evaluated very efficiently.

\begin{figure}
\centering
\includegraphics[width=0.5\textwidth]{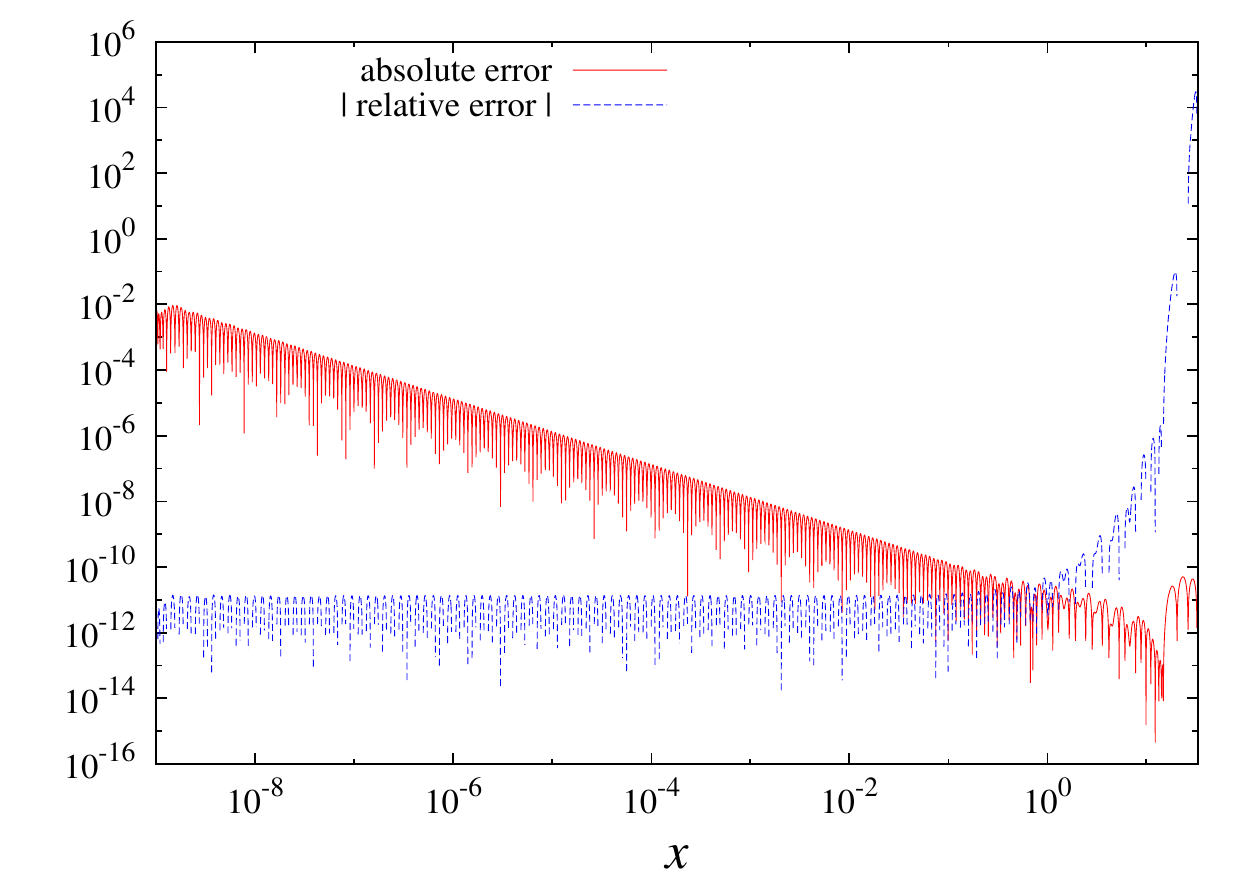}
\caption{Accuracy of the approximation  of the function $e^{-x}/x$ with 136 Gaussians used in the solution of the screened Poisson's equation for the case of free BC. The range of the independent variable $x$ is $[10^{-9},33]$. We plot both the absolute and the relative error because the latter is a better indicator close to the origin (where the fitted function takes on very large values), while the former is a reliable signature of the goodness of the fit towards the opposite end. Note that at $x=33$ the function $e^{-x}/x$ is already smaller than the machine precision.}
\label{fig:Gaussian_fit_Yukawa}
\end{figure}

We proceed by approximating the Green's function as
\begin{equation}\label{tensprod}
 G(\mu;r) \simeq \sum_k\omega_k(\mu) e^{-\alpha_k(\mu) r^2}\;,
\end{equation}
where  $\alpha_k(\mu)$ and $\omega_k(\mu)$ are determined so as to minimize the error on a given range of $r$. More specifically, we found out the following approximation,
\be\label{eq:Gaussian_fit1}
\frac{e^{-x}}{x}\simeq \sum_{k=1}^{136} \bar\omega_k e^{-\bar \alpha_k x^2},
\ee
with satisfactory accuracy for  any $x \in [10^{-9}, 33]$ (see Fig.\ \ref{fig:Gaussian_fit_Yukawa}).  The best fit was performed using the Levenberg-Marquardt algorithm \cite{Levenberg,Marquardt} with 136 Gaussian functions, the $\alpha_k$-values ranging between $10^{-3}$ and $10^{19}$. The actual Green's function can therefore be written as in Eq.\ \eqref{eq:Gaussian_fit1} with 
\be
\omega_k(\mu) = \frac{\mu \bar \omega_k}{4\pi},\qquad \alpha_k(\mu) = \mu^2 \bar \alpha_k.
\ee
In the unscreened case ($\mu_0 = 0$) we use the Gaussian fit of the $1/r$ function which was already proposed in Ref.\ \cite{genovese:074105} and that we recall as being affected by an error $\lesssim 10^{-8}$ for any $r\in [10^{-9},1]$.

The convolution kernel can be written as follows:
\be\label{eq:kernel}
K(j_x,j_y,j_z;\mu)  =  \sum_k \omega_k(\mu)\prod_{i \in \{x,y,z\}} I(\alpha_k(\mu) h_i^2; j_i)\,,
\ee
where
\be\label{eq:gaussian_ISF_convolution}
I(\alpha;j) \equiv \int dt\, e^{-\alpha (t-j)^2}\phi(t),
\ee
and can be computed by evaluating $(N_x+N_y+N_z)N_G$ 1D integrals - $N_G$ being the number of Gaussian functions -
hence at a much lower cost than Eq.\ \eqref{eq:kernel_FBC}, which would require the computation of $N_x N_y N_z$ 3D integrals, instead.

We point out that the numerical evaluation of Eq.\ \eqref{eq:gaussian_ISF_convolution} is performed using the same method described in Ref.\ \cite{genovese:074105}, which exploits the refinement relations fulfilled by the ISFs and yields an accuracy as high as the machine precision, even for the narrowest Gaussians. To this effect, ISFs show their superiority over other basis sets (cf.\ \emph{e.g.}\ the explicit method laid out in Ref.\ \cite{lee:224108}, where a Gaussian approximation similar to ours is carried out within a discrete variable representation approach).

\section{Wire-like Boundary Conditions}
\label{sec:WBC}
We now consider a system which is periodic along the $z$ direction (with period  equal to $L_z$) and isolated over the $xy$-plane. We can hence expand the continuous charge density distribution as a sum over its Fourier components along $z$:
\be\label{eq:wires_expansion}
\rho(x,y,z)=\sum_{p_z} e^{-2\pi i \frac{p_z}{L_z} z} \rho_{p_z}(x,y)\;,
\ee
noting that in this case the Fourier coefficients are solely functions of $x$ and $y$. After expanding the electrostatic potential in a similar manner, the screened Poisson's equation yields the following relation between the potential's reciprocal space components and those of the density:
\be\label{eq:screened_Poisson_WBC}
\left[\partial_x^2 + \partial_y^2 - \mu_0^2 - \mu_{p_z}^2 \right] V_{p_z}(x,y) =
-4\pi \rho_{p_z}(x,y)\;,
\ee
where $\mu_{p_z} \equiv 2 \pi p_z/{L_z}$. 
The symmetry of the problem suggests writing the Green's function of Eq.\ \eqref{eq:screened_Poisson_WBC} in cylindrical coordinates:
\be\label{eq:2D_Green_function}
\left[ \frac{\partial^2}{\partial r^2} + \frac{1}{r}\frac{\partial}{\partial r}- \mu^2 \right] G(\mu;r) = -\delta^{(2)}(r)\;,
\ee
where $r=\sqrt{x^2+y^2}$ and $\mu^2 = \mu^2_0 + \mu^2_{p_z}$. The solution of Eq.\ \eqref{eq:2D_Green_function} is given by
\be\label{formofgreen}
G(\mu; r)= \frac{1}{2\pi}
\begin{cases}
\textrm{K}_0(\mu r) & \mu > 0 \\
-\log (r) & \mu = 0
\end{cases}\,,
\ee
where $\textrm{K}_0$ is the zero-th order modified Bessel function of the second kind. We then express the 2D Fourier components of both density and potential in terms of ISFs, thereby completing the required steps towards the mixed plane-wave / ISF representation for the case investigated here:
\be\label{eq:mixed_representation}
\rho_{p_z}(x,y)=\sum_{j_x=0}^{N_x} \sum_{j_y=0}^{N_y} \;  \rho_{j_x,j_y;p_z} \,
\phi\left(\frac{x}{h_x}-j_x\right) \phi\left(\frac{y}{h_y}-j_y\right),
\ee
where $h_x$ and $h_y$ are the grid spacings along the non-periodic directions. Combining Eq.\ \eqref{eq:screened_Poisson_WBC} with Eq.\ \eqref{eq:wires_expansion} and Eq.\ \eqref{eq:mixed_representation} one obtains
\ba \label{zconvol}
V_{j_x,j_y;p_z} &=& 4 \pi h_x h_y \times  \\
&& \times \sum_{j'_x,j'_y} K(j_x-j'_x,j_y-j'_y;\mu) \rho_{j'_x,j'_y;p_z}\, ,\nonumber
\ea
where the kernel $K(j_x,j_y;\mu)$ is very similar to that in Eq.\ \eqref{eq:kernel_FBC}, except that in \eqref{zconvol} the integral is restricted to the non-periodic directions. Within the approximation \eqref{tensprod} the kernel elements can be evaluated as in Eq.\ \eqref{eq:kernel} with $i\in \{x,y\}$.

\begin{figure}
\centering
\includegraphics[width=0.5\textwidth]{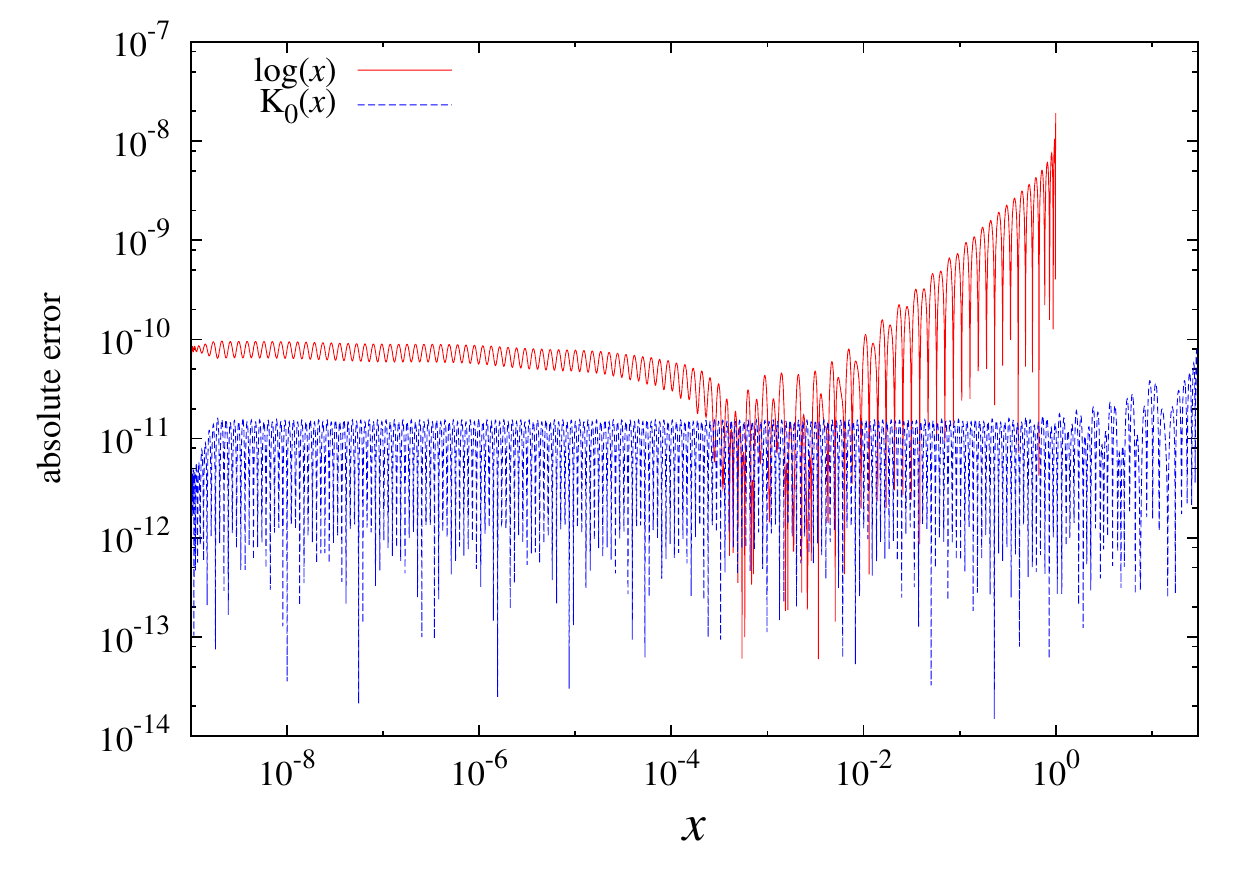}
\caption{Accuracy of the approximation of the Green's function with 144 Gaussians as used in the solution of the screened Poisson's equation for the case of wire-like BC. The range of the independent variable $x$ is $[10^{-9},30]$ for the function $\textrm{K}_0(x)$ and $[10^{-9},1]$ for $\log(x)$.}
\label{accuracy}
\end{figure}

Clearly, the accuracy of the whole method depends on the accuracy of the Gaussian approximation of the Green's function. Note that close to the origin the function $\textrm{K}_0(x)$ behaves like $\log(x)$, whereas it decreases exponentially for large arguments:
\begin{align}
\textrm{K}_0(x) \sim 
\begin{cases}
 -\log(x) & x \rightarrow 0\\
\frac{e^{-x}}{\sqrt{x}} & x\rightarrow \infty
\end{cases}\; .
\end{align}
We were able to approximate $\text{K}_0(x)$ as a sum of Gaussians (we denote the fitting parameters by $\{\bar \omega_k', \bar \alpha_k'\}$, cf.\  Eq.\ \eqref{eq:Gaussian_fit1}) with an accuracy better than $10^{-10}$ in the range $[10^{-9},30]$, where the upper bound for $x$ was chosen by realizing that the value attained by $\textrm{K}_0(x)$ at $x=30$ is $\simeq 2.1 \times 10^{-14}$, \emph{i.e.}\ already comparable with the machine precision. The function fitting was carried out using the same algorithm as before, but with 144 Gaussians, with $\bar \alpha'_k$-values ranging between $10^{-7}$ and $10^{20}$. We used the very same 144 Gaussians ($\bar \alpha_k'' \equiv\bar  \alpha_k'$) but different relative weights ($\bar \omega_k''$) also to approximate $\log(x)$, achieving an accuracy of $10^{-8}$ in the range $[10^{-9},1]$ (see Fig.\ \ref{accuracy}). 

\label{sec:num_results}

Therefore, for any $\mu > 0$ the convolution kernel is the same as in Eq.\ \eqref{eq:kernel} with $i\in \{x,y\}$, $\omega_k(\mu)  = \bar\omega_k'$ and $\alpha_k(\mu) = \bar \alpha_k' \,\mu^2$, whereas for $\mu = 0$ we exploit the scaling properties of the logarithm to adapt the best fit obtained for $x\in [0,1]$ to any $r \in [0,L]$. We write it explicitly for sake of clarity: 
\ba
 K(j_x,j_y;\mu = 0) &=& \log (L) + \\ && + \sum_k\bar \omega''_k I\left(\frac{\bar \alpha''_k}{L^2} h_x^2;j_x \right) I\left(\frac{\bar\alpha''_k}{L^2} h_y^2;j_y\right), \nonumber
\ea
where $L\equiv\sqrt{[h_x(N_x+2m)]^2 + [h_y(N_y+2m)]^2}$ is deliberately chosen so that $r/L<1$. In fact, $h_{\{x,y\}} N_{\{x,y\}}$ is the box size along ${\{x,y\}}$, \emph{i.e.}\ the maximum range covered by the charge distribution along each axis; $2m\,h_{\{x,y\}}$ is the extent of the ISF definition domain. The convolution of any charge distribution function with an ISF can  be non-zero at most over a rectangular domain of diagonal length $L$.

In order to find the solution for the electrostatic potential in real space, we first compute the Fourier coefficients of the density ($\rho_{j_x,j_y;p_z}$) through a 1D FFT along the periodic direction $z$. The corresponding quantities for the potential are then obtained by calculating the convolutions in Eq.\ \eqref{zconvol} via a zero-padded FFT procedure \cite{Goedecker1993294}. Finally, the potential is transformed back into real space along the $z$ direction. Let us notice that real-to-complex FFTs can be used instead of complex-to-complex FFTs, since all the quantities are real and the kernel is symmetric.

\section{Surface-like Boundary Conditions}
\label{sec:SBC}
While referring the reader to Ref.\ \cite{genovese:054704} for a more extensive description of our treatment of surface-like BC, in the following we just discuss our methodological improvements . In particular, we are now able to allow also for the screening and for monoclinic lattices, simply by redefining the $\mu^2$ in the equation relating the 2D Fourier components of density and potential,
\be
\left(\frac{\partial^2}{\partial z^2} - \mu^2\right)V_{p_x,p_y}(z) = -4\pi\,\rho_{p_x,p_y}(z),
\ee
as follows:
\be
\mu^2 = \mu^2_0+4\pi^2 \sum_{i,j}g^{ij}\frac{p_i}{L_i}\frac{p_j}{L_j}\,,
\ee
where $i,j\in \{x,y \}$ (\emph{i.e.}\ the periodic directions with period equal to $L_x$, $L_y$) and $g^{ij}$ is the 2D contra-variant metric tensor,
\be\label{eq:contravariant_metric}
g^{ij} = \frac{1}{\sin^2 \alpha}\left[
\begin{array}{cc}
1 & -\cos\alpha  \\
-\cos\alpha & 1
\end{array}
 \right],
\ee
$\alpha$ being the angle between the $\hat x$ and $\hat y$ unit vectors.

We point out that in the case of surface-like BC there is no Gaussian approximation of the Green's function involved in the procedure. Consequently, in this case the accuracy of the method is limited only by the machine precision.

\section{Numerical results}

In order to measure the accuracy of our method, we used several test charge distributions for which the Poisson's equation is exactly solvable, and compared the approximate numerical solution against the exact one. In the following the accuracy is given in terms of the infinity norm:
\be
||\mbox{err}||_\infty = \max_{j_x,j_y,j_z}|V_{j_x,j_y,j_z}-V_{j_x,j_y,j_z}^{\textrm{(exact)}}|.
\ee
Without loss of generality, all tests were run on a cubic  simulation box ($L_x = L_y = L_z\equiv L$), with a uniform grid and an equal number of points along each direction.

\begin{figure}
\centering
\includegraphics[width=0.5\textwidth]{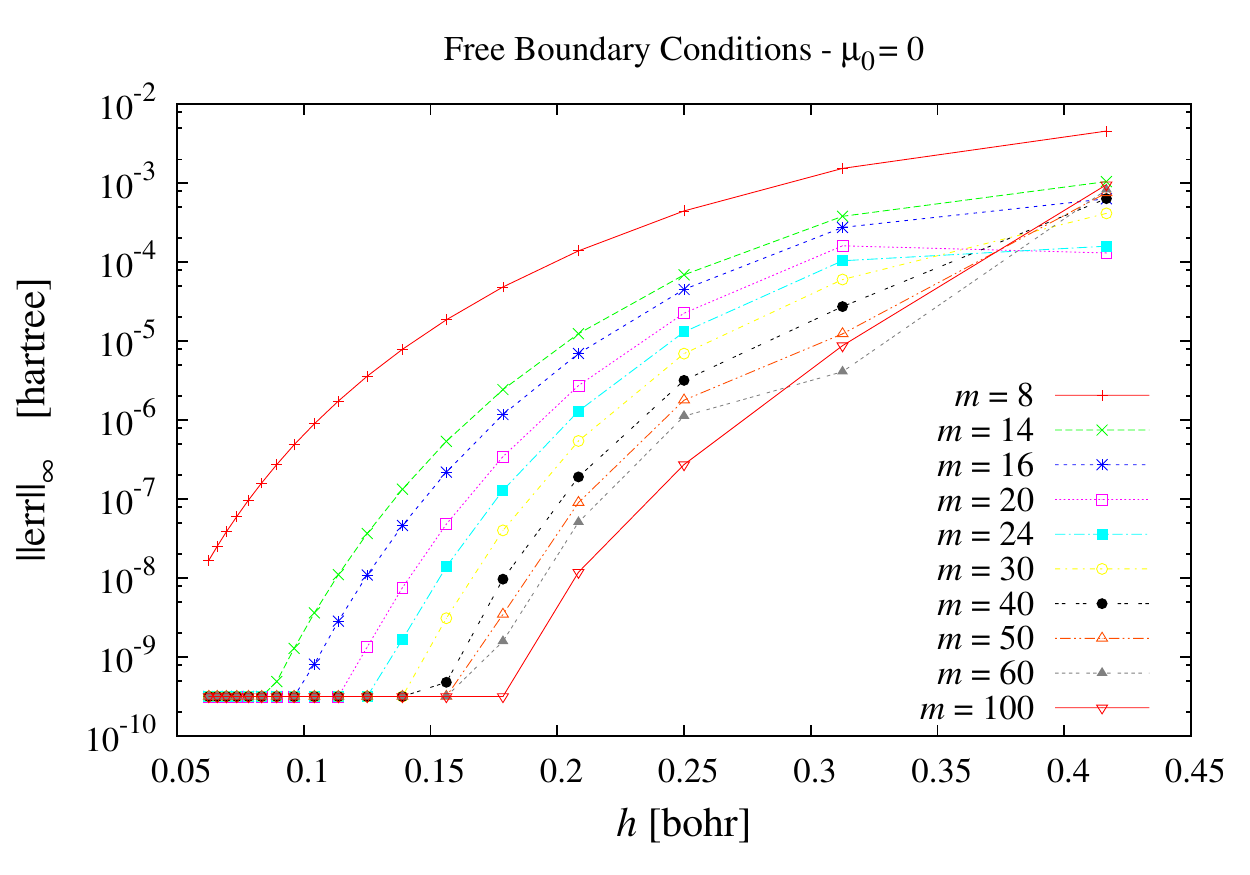}
\caption{Accuracy test for the case of free/isolated boundary conditions in the absence of screening ($m$ is the order of the ISF, $h$ the grid spacing).
}
\label{fig:accuracy_FBC}
\end{figure}

\begin{figure}
\centering
\includegraphics[width=0.5\textwidth]{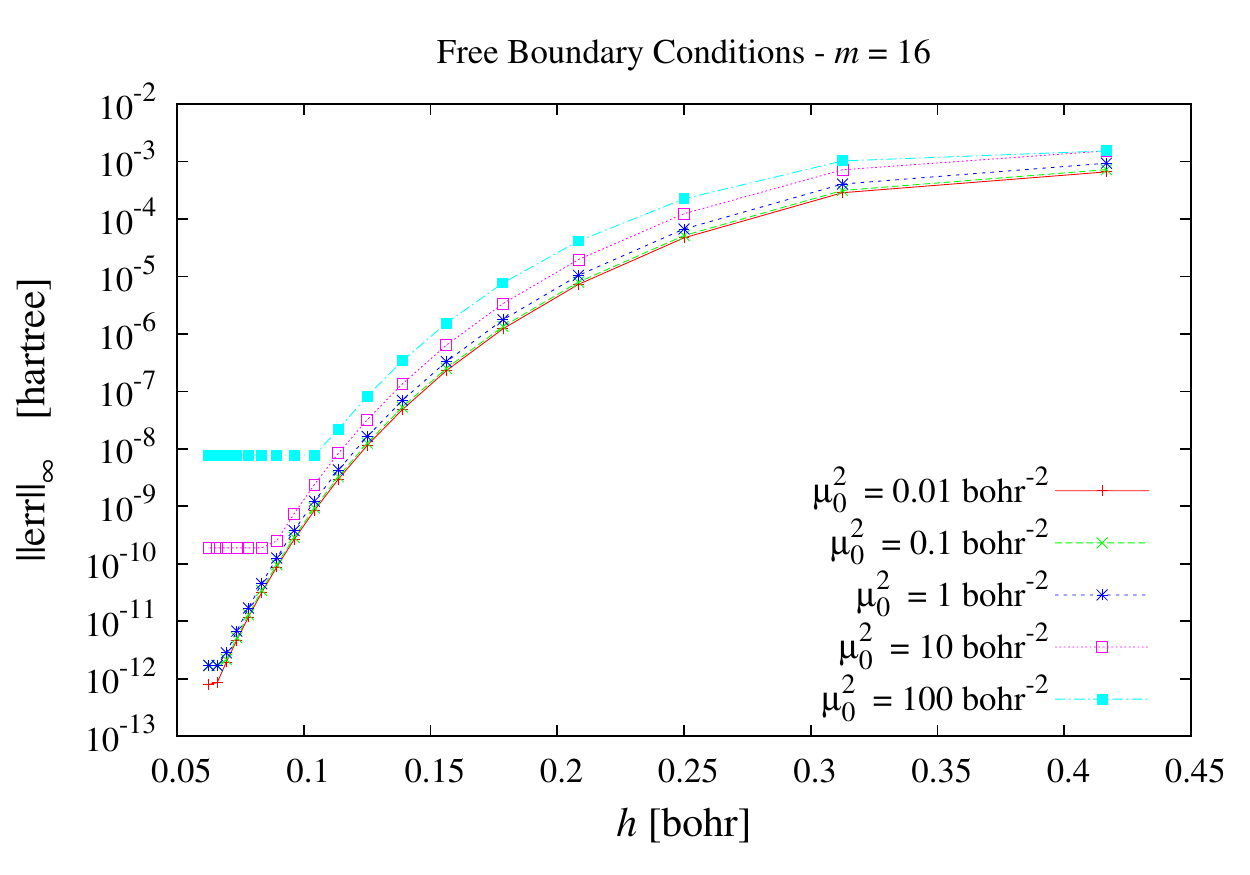}
\caption{Accuracy test for the case of free/isolated boundary conditions in the presence of screening ($m=16$ is the order of the ISF, $h$ the grid spacing).
}
\label{fig:accuracy_FBC_2}
\end{figure}

\begin{figure}
{\includegraphics[width=0.5\textwidth]{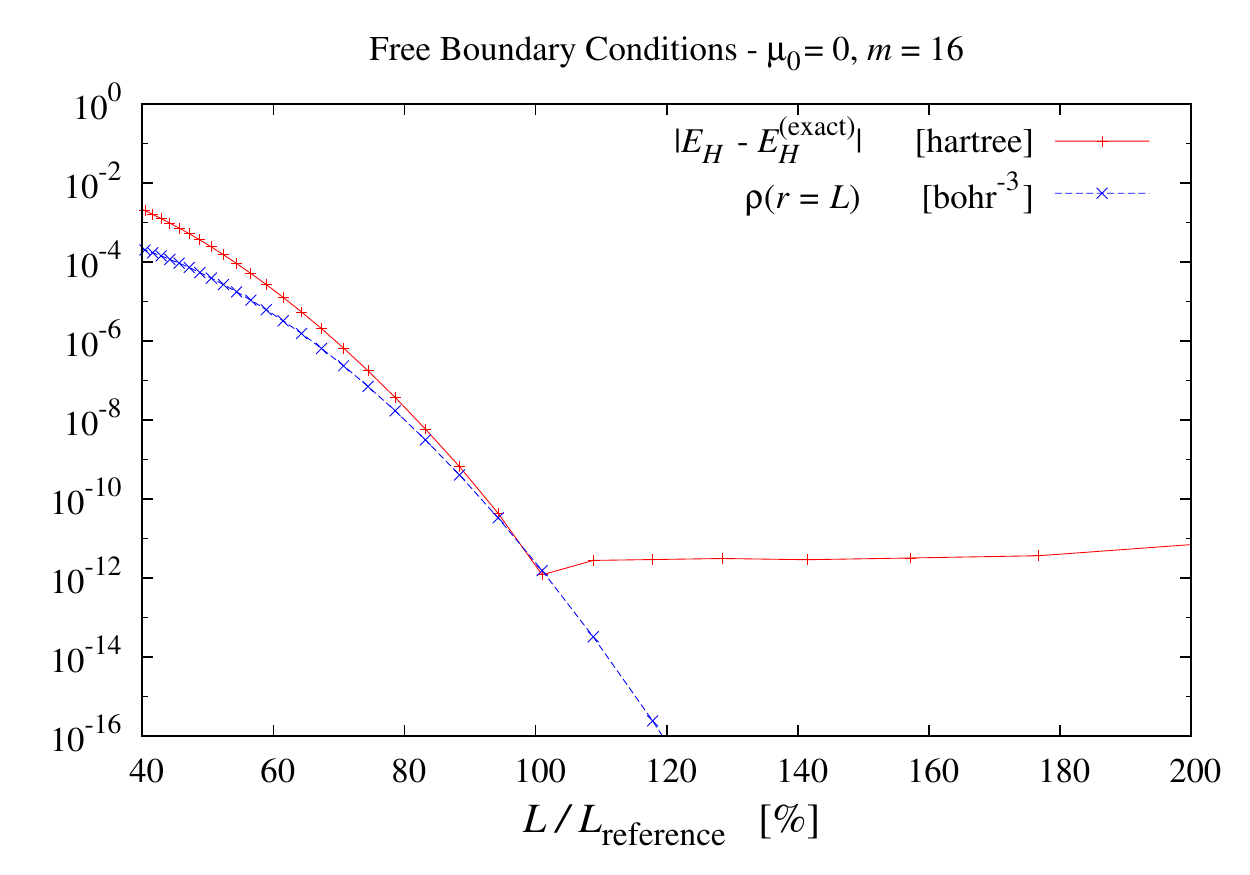}}

\caption{Influence of the (cubic) simulation box size on the accuracy of the Hartree energy. $L$ stands for the box size, whereas $L_{\mbox{ref.}}$ stands for the box size for which $\rho(r=L) =2.21\times 10^{-12} \;\mbox{bohr}^{-3}$.}
\label{fig:accuracy_FBC_3}
\end{figure}

In particular, in the case of free BC we used a Gaussian density distribution, $\rho(r) = A\, \exp\left[-r^2/(2\sigma^2)\right]$. As our procedure relies on a convolution - see Eq.\ \eqref{eq:convolution_integral} - from which no divergence can arise (as long as $\rho(\vec{r})$ is regular), we expect that the numerical solution is regular everywhere.  The latter statement offers an unambiguous prescription for fixing the integration constants of the analytic exact solution. The latter is eventually found to be as follows:
\ba\label{eq:exact_V_FBC}
&& V(r) = A\,(\sqrt{2\pi}\sigma)^3\frac{e^{-\mu_0 r +\mu_0^2\sigma^2/2}}{2r}\times\\
&& \times\left[\mbox{erfc}\left(-\frac{r}{\sqrt{2}\sigma}+\frac{\mu_0\sigma}{\sqrt{2}}\right) -e^{2\mu_0 r}\mbox{erfc}\left(\frac{r}{\sqrt{2}\sigma}+\frac{\mu_0\sigma}{\sqrt{2}}\right) \right]. \nonumber
\ea
The results of our tests - with $A$  such that $V(r\rightarrow 0) =1$ - are reported in Fig.\ \ref{fig:accuracy_FBC} for all the ISF supported in our code but  no screening and in Fig.\ \ref{fig:accuracy_FBC_2} for 16th order ISF and selected values of $\mu_0^2$ covering four orders of magnitude. We point out that, owing to the chosen value of $\sigma$, the value of the density on the simulation box faces is $\rho(r=L) =2.21\times 10^{-12} \;\mbox{bohr}^{-3}$, hence smaller than our solver's accuracy. In order to study the influence of the box size on accuracy, we performed several runs with different box sizes, computed the Hartree energy, and compared the result to the exact Hartree energy corresponding to an infinite box size and unitary monopole ($q\equiv \int \dd^3 \vec{r} \rho(\vec{r}) = 1$):
\be
E_H^{(\mbox{exact})} \equiv \frac{1}{2}\, \int \dd^3 \vec{r}\;\rho(\vec{r}) V(\vec{r}) =\frac{1}{2\sigma \sqrt{\pi}}.
\ee
Throughout this second set of tests the value of $A$ was chosen so as to keep $q = 1$. The results are reported in Fig.\ \ref{fig:accuracy_FBC_3}, together with the value of the charge density distribution at the box faces. We can observe that there is absolutely no need to enlarge the box size beyond the reference size, as the latter is already large enough to capture all the features within the reach of our solver. On the other hand, accuracy decreases on reducing the box size, as the ideal free BC $\rho({r\rightarrow L})\rightarrow 0$ becomes increasingly violated. Nevertheless, the attained accuracy remains interesting over a broad range of box sizes even smaller than the reference one, at variance with plane-wave methods which would fail our test, firstly because of the presence of a non-zero monopole and secondly because of the aliasing due to insufficiently large box. 

We remark that, as opposed to what claimed in Ref.\ \cite{hine:204103} in relation to Refs.\ \cite{genovese:054704,genovese:074105}  (where the method deployed here was first proposed), our solver is actually reliable without any resort to the so-called ``minimum image convention'', namely without rendering the input density charge distribution periodic within the simulation box.


\begin{figure}
\centering
\includegraphics[width=0.5\textwidth]{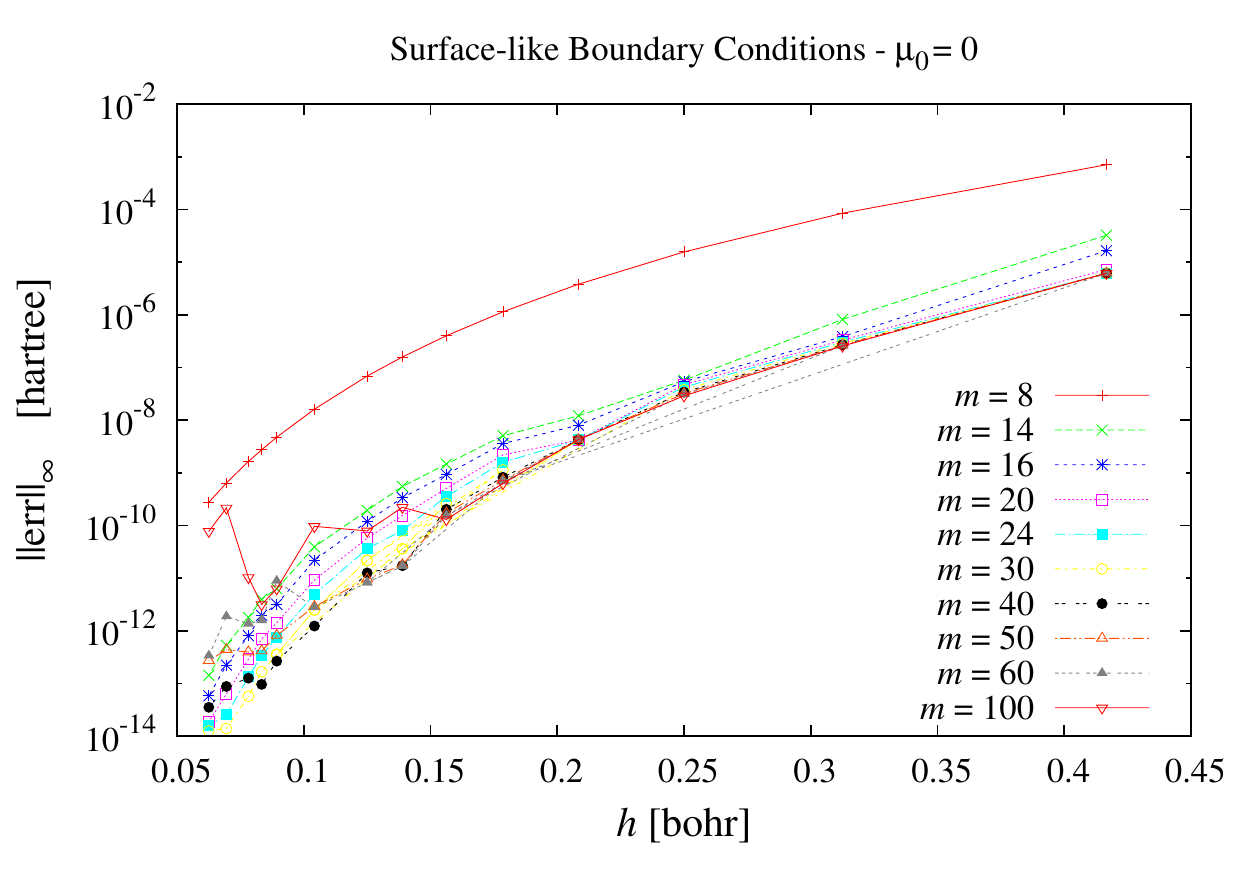}
\caption{Accuracy test for the case of surface-like boundary conditions in the absence of screening ($m$ is the order of the ISF, $h$ the grid spacing).}
\label{fig:accuracy_SBC}
\end{figure}

\begin{figure}
\centering
\includegraphics[width=0.5\textwidth]{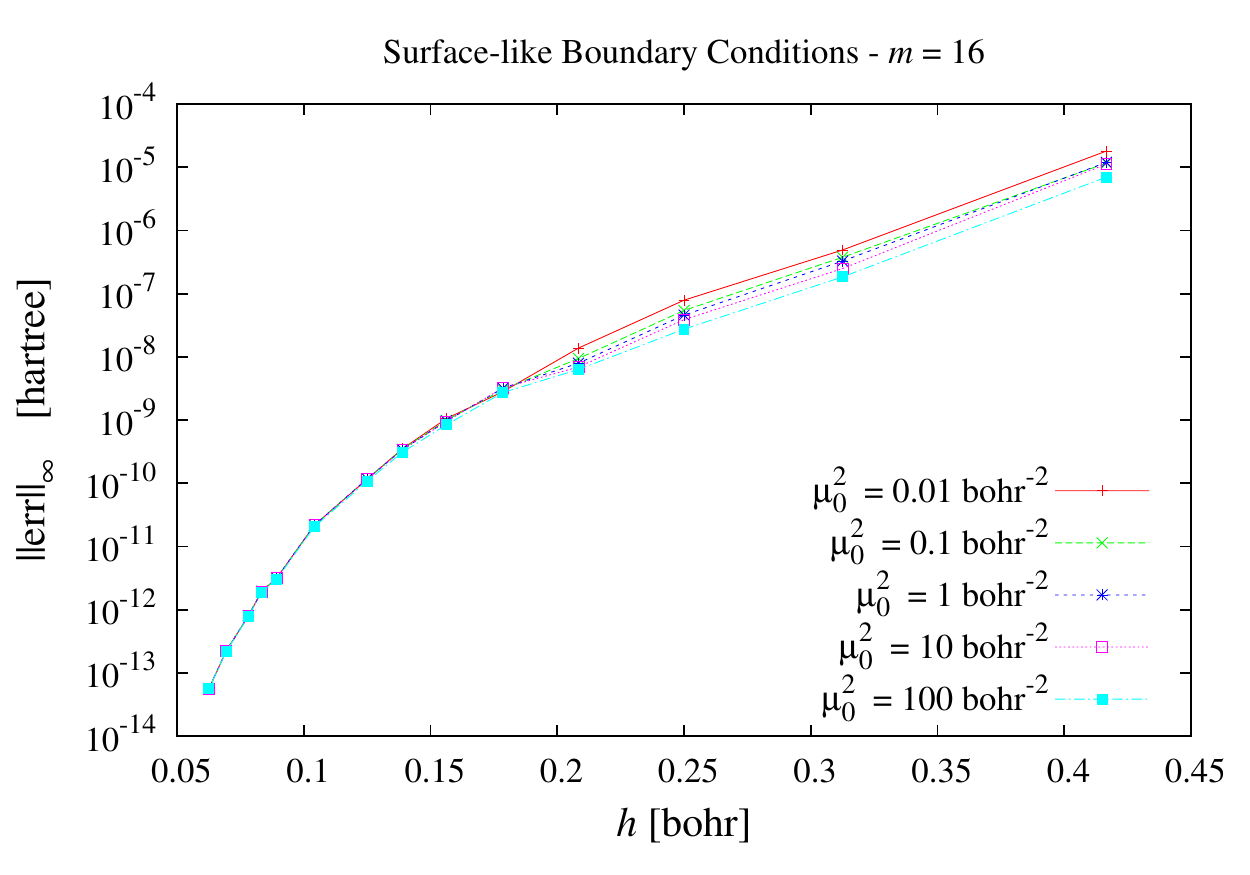}
\caption{Accuracy test for the case of surface-like boundary conditions in the presence of screening ($m=16$ is the order of the ISF, $h$ the grid spacing).}
\label{fig:accuracy_SBC_2}
\end{figure}

\begin{figure}
\centering
\includegraphics[width=0.5\textwidth]{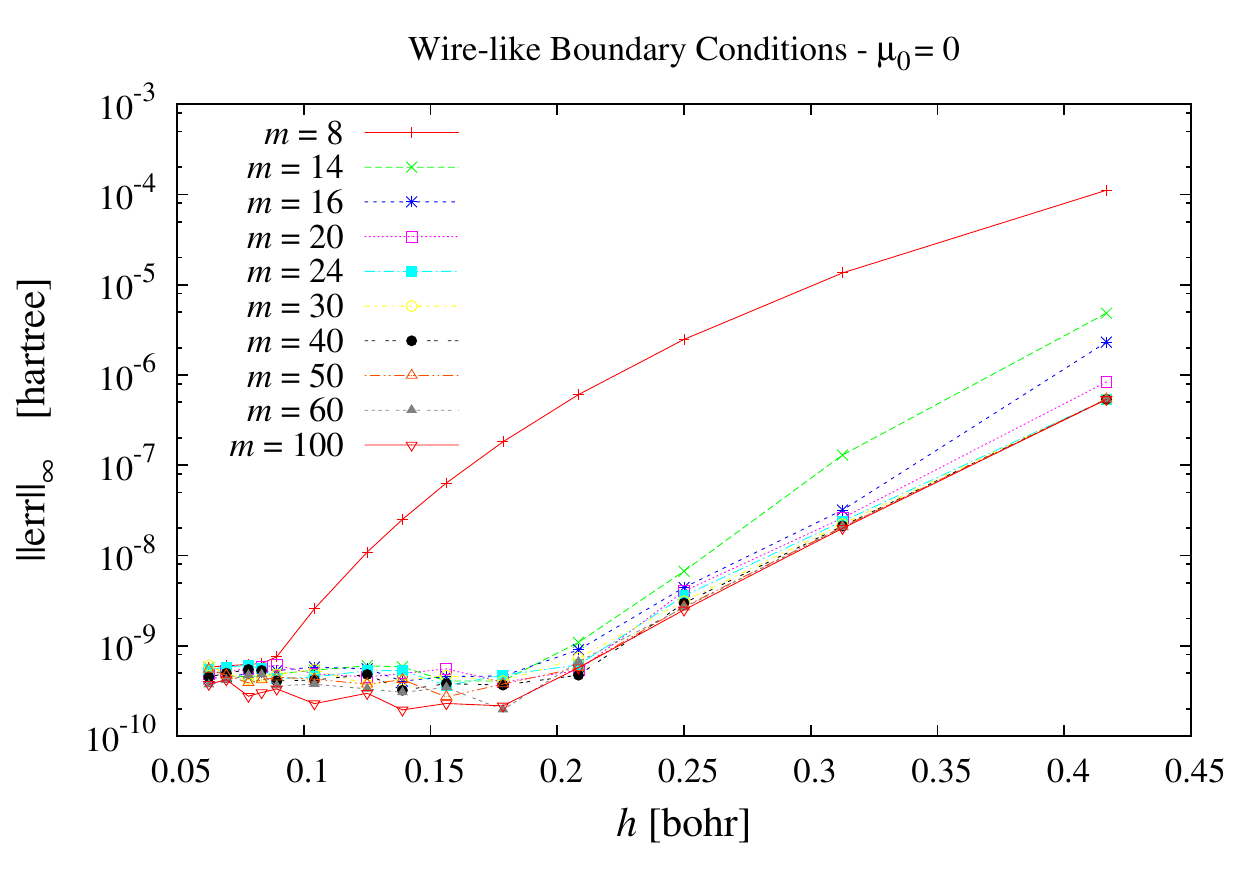}
\caption{Accuracy test for the case of wire-like boundary conditions in the absence of screening ($m$ is the order of the ISF, $h$ the grid spacing).}
\label{fig:accuracy_WBC}
\end{figure}

\begin{figure}
\centering
\includegraphics[width=0.5\textwidth]{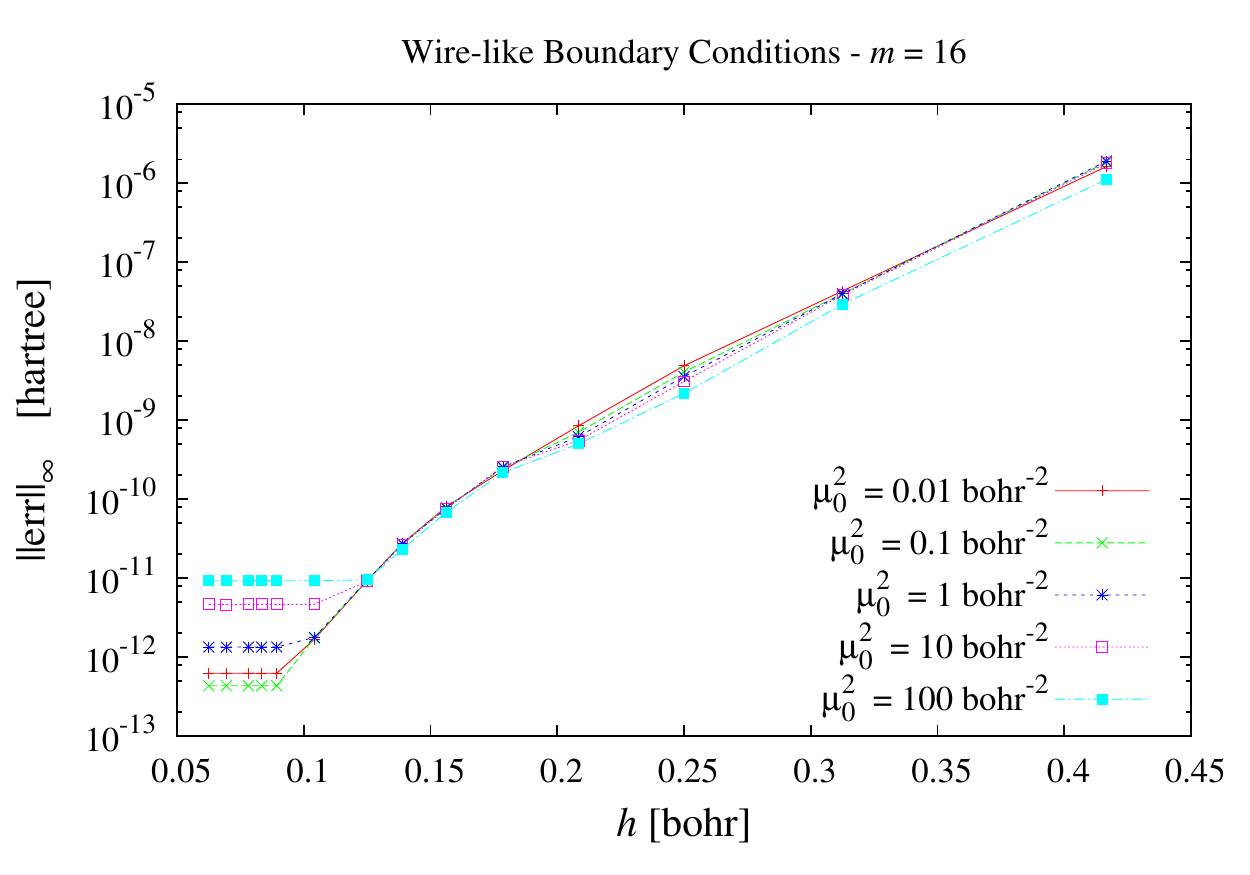}
\caption{Accuracy test for the case of wire-like boundary conditions in the presence of screening ($m=16$ is the order of the ISF, $h$ the grid spacing).}
\label{fig:accuracy_WBC_2}
\end{figure}

In the case of surface-like and wire-like BC, we consider a charge density distribution obtained by applying explicitly the screened Poisson's operator on an exact potential written as
\be
V(x,y,z) = -4\pi \prod_{i\in\{x,y,z\}} f_{\{P,I\}}(i;L_i)\;,
\ee
where each of the \emph{ad hoc} functions $f$'s entering the product is either periodic,
\ba 
f_P(x;L) &=& \exp \left [\cos \left ( 2\pi \frac {x-L/2}{L}\right )\right]\;,
\ea
or localized,
\ba
f_I(x;L) &=& \exp \left [-50\left ( \frac {x-L/2}{L}\right )^2\right ]\times\nonumber
\\ && \times \exp \left [-\tan ^2\left (\pi \frac {x-L/2}{L}\right )\right ]\;,
\ea
depending on the intended BC. Results are shown in Figs.\ \ref{fig:accuracy_SBC}, \ref{fig:accuracy_SBC_2},  \ref{fig:accuracy_WBC}, \ref{fig:accuracy_WBC_2} and indicate a very good overall convergence rate.

\begin {figure}
\centering 
\includegraphics [width=0.5\textwidth ]{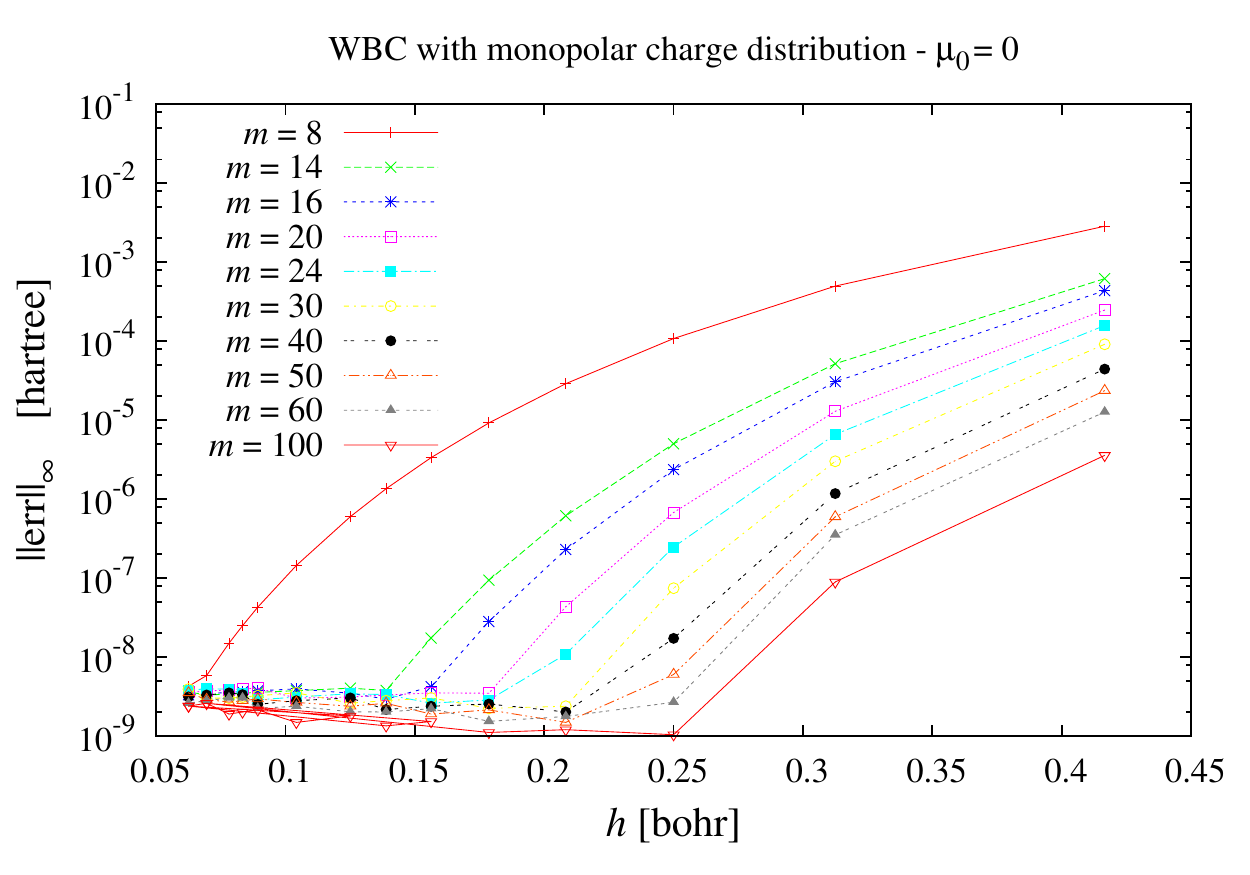}
\caption {Accuracy test for the case of wire-like boundary conditions (WBC) with monopolar charge density distribution ($m$ is the order of the ISF, $h$ the grid spacing).}
\label{fig:accuracy_WBC_monopolar}
\end {figure}

\begin {figure}
\centering 
\includegraphics [width=0.5\textwidth ]{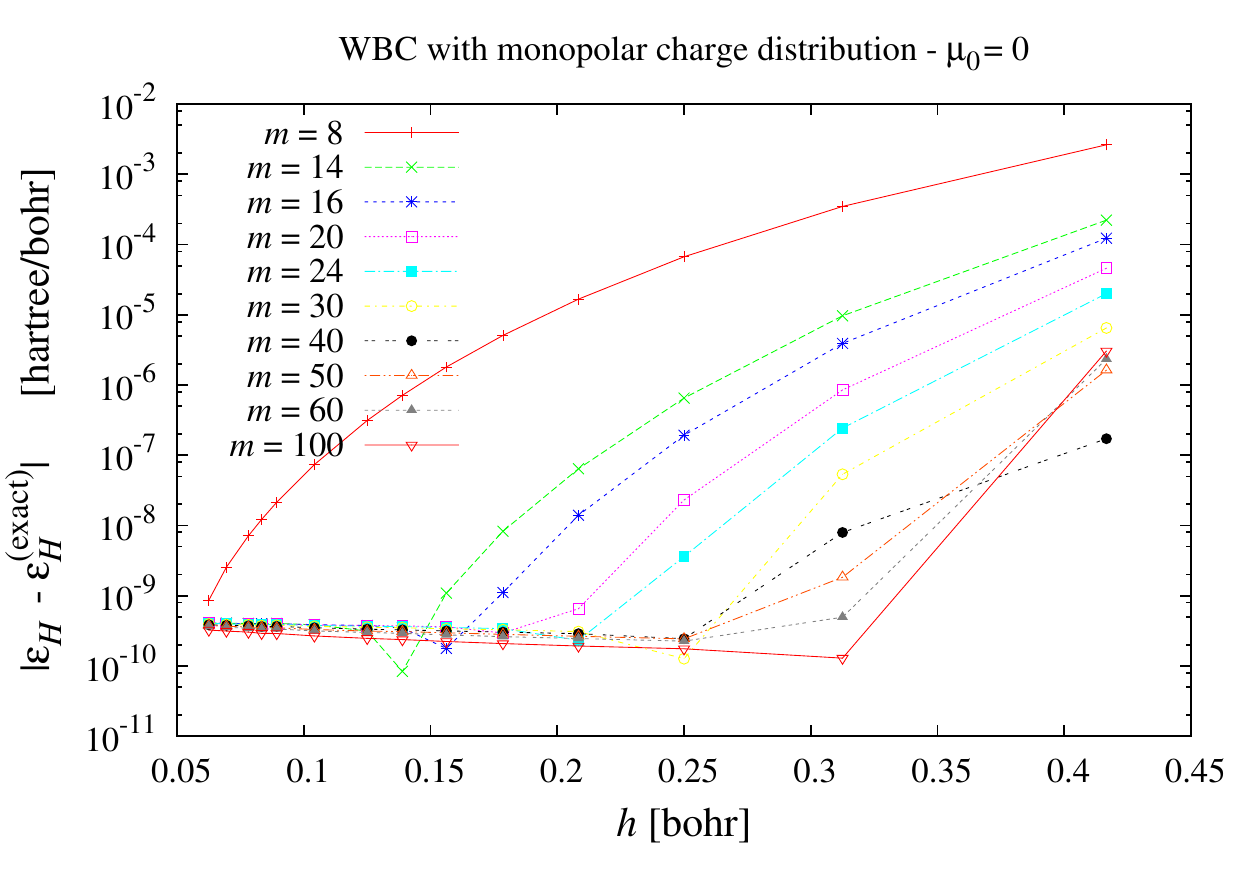}
\caption {Accuracy in the computation of the Hartree linear energy density - see Eq.\ \eqref{eq:Hartree_linear_density} -  in the case of wire-like boundary conditions (WBC) with monopolar charge density distribution ($m$ is the order of the ISF, $h$ the grid spacing). In our setup $\varepsilon_H^{\mbox{(exact)}} = 7.1211128333623614\;\mbox{hartree/bohr}$.}
\label{fig:accuracy_WBC_monopolar_2}
\end {figure}

\begin {figure}
\centering 
{\includegraphics[width=0.5\textwidth]{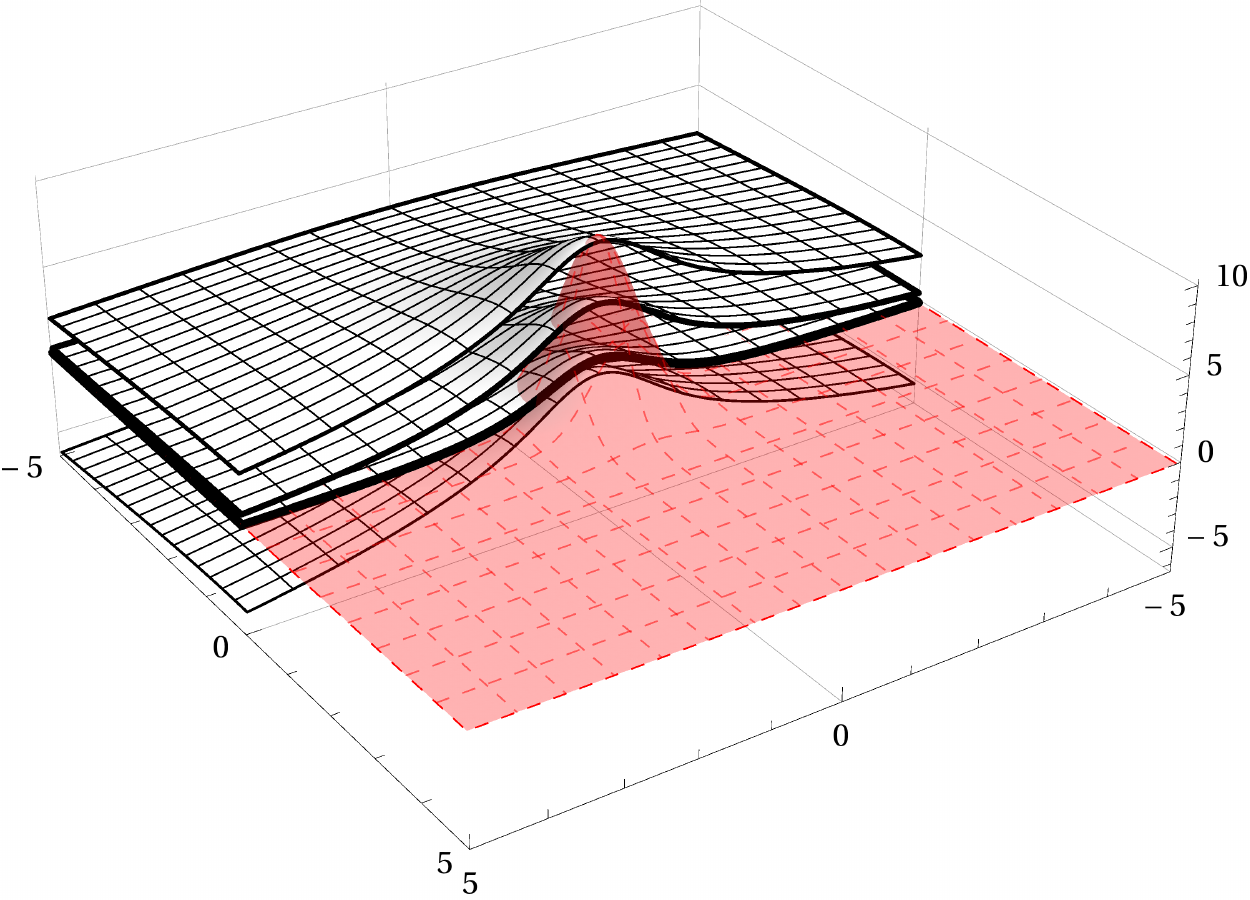}}
\caption{
Gaussian density charge distribution (red, dashed mesh) as a function of the isolated directions ($x$, $y$ in our notation) and the corresponding electrostatic potential (black, solid mesh) evaluated for different values of the screening, namely $\mu_0^2 = \{0, 0.01, 0.10,1\}\,\mbox{bohr}^{-2}$ upon increasing boundary thickness. The charge density is implicitly periodic along $z$ (wire-like BC). The amplitude of the plotted density charge distribution is multiplied by a factor $10$ with respect to the actual value in order to improve the readability of the picture. Only half of the solution is drawn to highlight its profile.}
\label{fig:screened_monopole}
\end {figure}

In the case of wire-like BC, we furthered our tests by choosing a 2D Gaussian charge  distribution,
\be\label{eq:monopolar_density_distr}
\rho(r,z)\equiv \rho(r) = e^{-k r^2},\quad r=\sqrt{x^2+y^2},
\ee 
where the charge density along $z$ is implicitly set to unity. The corresponding exact potential to be used as reference is
\be\label{eq:monopolar_potential}
V(r) = 4\pi \left[\frac{\mbox{Ei}(-kr^2)-\log(r^2)}{4k}\right],
\ee
where $\mbox{Ei}(x)$ is the exponential integral function.
On deriving Eq.\ \eqref{eq:monopolar_potential}, integration constants were fixed unambiguously by following the same criteria which led to Eq.\ \eqref {eq:exact_V_FBC}. In particular, one integration constant guarantees the regularity of the potential at the origin,
\be
\lim_{r\rightarrow 0} V(r) = \frac{\pi}{k}\left[ \gamma_E+\log(k)\right] \;,
\ee
where $\gamma_E$ is the Euler-Mascheroni constant, while the second integration constant (an additive constant) is set to zero so that $V(r)\sim -\log(r)$ as $r\rightarrow \infty$, in accordance with the behavior of the Green's function. This test case is relevant to our studies for a two-fold reason. Firstly, it is typically out-of-reach for plane-wave methods, since the density distribution in Eq.\  \eqref{eq:monopolar_density_distr} exhibits a non-zero monopole.
Secondly, it allows to probe the goodness of the Gaussian approximation of the $\log(x)$ function involved in Eq.\ \eqref{formofgreen} within the approximation \eqref{tensprod}. The charge distribution being constant along $z$, the only non-trivial term in the Fourier expansion along $z$ is the zero-mode ($\mu_{p_z} = 0$) and, in case the screening is absent ($\mu_0 = 0$), only the $\log$-branch of the Green's function \eqref{formofgreen} plays a role in the computation. We have also analyzed the rate of convergence towards the exact Hartree linear energy density,
\ba\label{eq:Hartree_linear_density}
\varepsilon_H^{(\mbox{exact})} &\equiv& \frac{1}{2}\, \int \dd^2 \vec{r}\;\rho(\vec{r},z) V(\vec{r},z) = \\
&=&\frac{\pi^2}{2k^2}\left(\gamma_E + \log\frac{k}{2}\right),
\ea
so as to have a further confirmation that the great accuracy in the numerically evaluated electrostatic potential ensures the reliable computation of derived physical quantities. The results are shown in Figs.\ \ref{fig:accuracy_WBC_monopolar}-\ref{fig:accuracy_WBC_monopolar_2} and are definitely good. We also include in Fig.\ \ref{fig:screened_monopole} a plot of the charge density distribution (magnified by a factor 10) and the corresponding potentials obtained at different $\mu_0$'s. We observe that for $\mu_0=0$ the potential does not fall to zero for increasing $r$, whereas it tends to be more and more localized around the origin as $\mu_0$ increases, as expected.

\begin {figure}
\centering 
{\includegraphics[width=.5\textwidth]{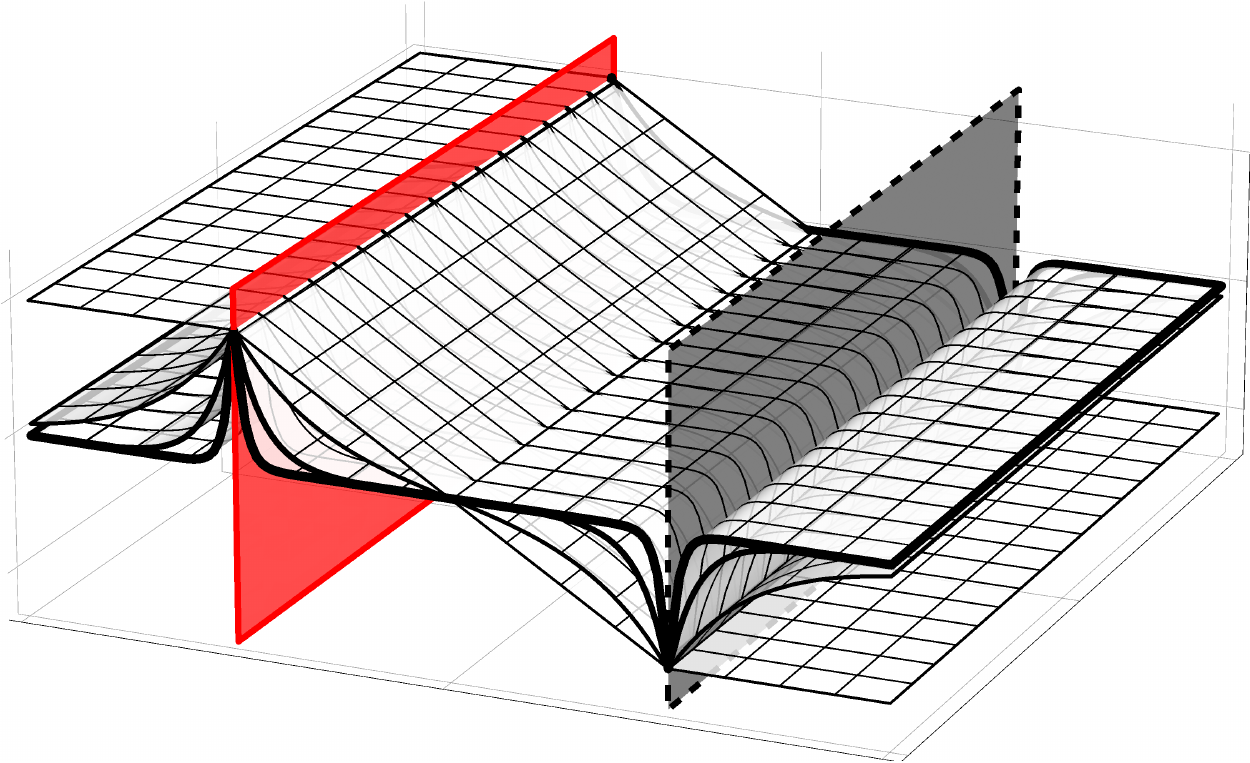}}
\caption{
Electrostatic potential generated by a pair of planar charge distributions of opposite sign (positive in red/solid; negative in black/dashed) modelling a planar capacitor unlimited in the periodic directions. The piecewise planar behavior corresponds to the case with no screening ($\mu_0 = 0$), whereas the other solutions are obtained with $\mu_0^2=\{1,10,100\}\,\mbox{bohr}^{-2}$ upon increasing the boundary thickness. The potential is more and more localized around the capacitor's plates and falls rapidly to zero as $\mu_0$ is increased. Each curve is normalised to one to improve readability.}
\label{fig:plane_capacitor}
\end {figure}

\begin{figure}
\centering 
{\includegraphics[width=0.5\textwidth]{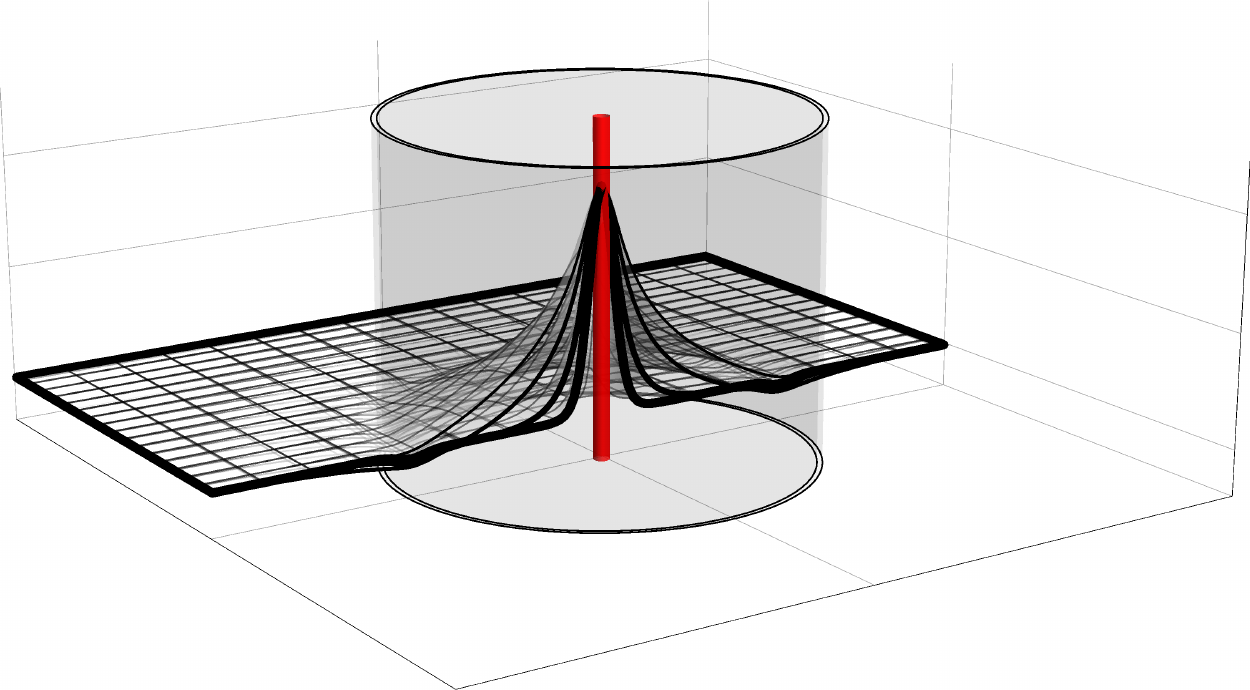}}
\caption{Electrostatic potential generated by a cylindrical capacitor, periodic in the vertical direction. The different solutions correspond to $\mu_0^2=\{0,1,10,100\}\,\mbox{bohr}^{-2}$ upon increasing the boundary thickness. Each curve is normalised to one for sake of readability. Only half of the solution is drawn so as to highlight the potential profile.}
\label{fig:cylindrical_capacitor}
\end {figure}

Having checked that our solver performs quite well in all the above mentioned cases, it seemed tempting to further probe its capabilities with some other charge density distributions. In particular, we modelled a planar capacitor (hence involving surface-like BC), and a cylindrical capacitor (wire-like BC). In the latter case, the input charge distribution was mimicked by a positive 2D Gaussian distribution sharply peaked at the origin, together with a ring of negative Gaussian distributions centered at $r = L/4$. The relative amplitudes of the central and of the peripheral Gaussians were chosen to yield zero total charge. The solutions obtained are shown in Figs.\ \ref{fig:plane_capacitor}-\ref{fig:cylindrical_capacitor}, are definitely consistent with the intuitively expected behavior.

\section{Conclusion}
We have presented a numerical method for the solution of Poisson's equation which can tackle any type of periodicity, the presence of screening, non-orthorhombic geometries and charged systems, showing that convergence to highly accurate results is attained with no adjustable  parameter other than the (unavoidable) grid spacing. 

The charge density distribution and the electrostatic potential are both expressed in terms of plane waves along the periodic directions, and in terms of  interpolating scaling functions along the non-periodic directions. The latter representation proves to be very handy, in that the expansion coefficients of any continuous quantity are simply given by its values on a uniform grid. Moreover, $m$-th order interpolating scaling functions preserve the matching between the moments built upon the discretized quantity with those of the continuous one. This occurrence is particularly important in electrostatics, because of the interest in resolving the multipolar features of the electrostatic potential and other derived quantities.

In our approach, the solution is obtained via the Green's function method. The (in principle) most expensive operation, namely the convolution of the Green's function with the input charge density distribution, becomes affordable by making use of highly-optimized  $\mathcal{O}(N\log N)$ FFT routines (zero-padded along the isolated directions), while the evaluation of the convolution kernel is carried out separately along the three spatial directions by approximating the Green's function as a sum of Gaussian functions.

Owing to the above mentioned advantages, our solver is  suitable for intensive computer simulations of electronic structure and molecular dynamics, where the Poisson's equation has to be solved several times and it is important to limit the growth and propagation of numerical errors as much as possible, especially because other physical quantities are computed starting from the solution of the Poisson's equation. There are other contexts in computational physics and chemistry in which relevant quantities are obtainable as convolutions involving the same Green's functions found in electrostatics. This is the case, for instance, of the exact exchange term within the generalizations of Kohn-Sham DFT employing orbital-dependent or hybrid functionals. Actually, our methodology features a level of generality which allows to address also problems well beyond electrostatics. 

As a possible outlook, we are working towards enabling the computation of range-separated hybrid functionals, where the Coulomb potential is split into a long- and a short-range component. In the basic range-separated approach the Coulomb interaction is written as $1/r =\left[ \mbox{erf}(r/r_0)+\mbox{erfc}(r/r_0)\right]/r$, $r_0$ being an adjustable length scale, although other options have been put forward (in which, for instance, the long- and short-range parts can be weighted differently). We are thus planning to model the range-separated Coulomb interaction within our framework.

Our solver is already designed for taking full advantage of multi-core CPUs and GPUs, and is currently integrated in \textsc{BigDFT}, the sources of which are freely downloadable from \url{http://inac.cea.fr/L_Sim/BigDFT/}. The release of a stand-alone package is also envisaged for the near future. The details on the GPU acceleration and the performance of the solver in the context of massively parallel electronic structure computations will be described in a forthcoming paper.


\acknowledgements
The authors thank Thierry Deutsch for valuable suggestions on the manuscript, and Claudio Ferrero for the critical proofreading.
A.C. acknowledges the financial support of the French National Research Agency in the frame of the ``\textsc{NEWCASTLE}'' project.

\end{document}